\documentclass[pra,superscriptaddress,showpacs,floatfix]{revtex4}
\usepackage{graphicx}
\usepackage{epstopdf}
\usepackage{amsmath}
\usepackage{bm}
\usepackage{color}

\newcommand{\red}[1]{{\color{red}#1}}

\begin{document}
\title{Full counting statistics and phase diagram of a dissipative Rydberg gas}
\author{N. Malossi$^{1,2}$, M.M. Valado$^{1,2}$, S. Scotto$^2$, P. Huillery$^3$, P. Pillet$^3$, D. Ciampini$^{1,2,4}$, E. Arimondo$^{1,2,4}$        	     \& O. Morsch$^{1,2}$}
\affiliation{$^1$INO-CNR, Via G. Moruzzi 1, 56124 Pisa, Italy\\ $^2$Dipartimento di Fisica `E. Fermi', Universit\`a di
Pisa, Largo Pontecorvo 3, 56127 Pisa, Italy\\$^3$ Laboratoire Aim\'{e} Cotton, CNRS, Univ Paris-Sud 11, ENS-Cachan, Campus d'Orsay Bat. 505, 91405 Orsay, France \\$^4$CNISM UdR Dipartimento di Fisica `E. Fermi', Universit\`a di Pisa, Largo Pontecorvo 3, 56127 Pisa, Italy}

\begin{abstract}
Ultra-cold gases excited to strongly interacting Rydberg states are a promising system for quantum simulations of many-body systems \cite{verstraete2009,Weimer2010}. For off-resonant excitation of such systems in the dissipative regime, highly correlated many-body states exhibiting, among other characteristics, intermittency and multi-modal counting distributions are expected to be created \cite{Ates2012,Lee2011,Lee2012,Hu2013}. So far, experiments with Rydberg atoms have been carried out in the resonant, non-dissipative regime \cite{Comparat2010}. Here we realize a dissipative gas of rubidium Rydberg atoms and measure its full counting statistics for both resonant and off-resonant excitation. We find strongly bimodal counting distributions in the off-resonant regime that are compatible with intermittency due to the coexistence of dynamical phases. Moreover, we measure the phase diagram of the system and find good agreement with recent theoretical predictions \cite{Hu2013,Ates2012,Lee2012}. Our results pave the way towards detailed studies of many-body effects in Rydberg gases.

\end{abstract}


\maketitle

Ultra-cold atoms excited to high-lying Rydberg states have been shown in recent years to be promising candidates for experimental implementations of quantum computation and quantum simulation \cite{Saffman2010}. The strong and controllable interactions between Rydberg atoms mean that fast two-qubit quantum gates and models of many-body Hamiltonians can, in principle, be efficiently realized. As a further important ingredient, dissipation has been shown to lead to novel phases and to facilitate, under certain conditions, the creation of pure and coherent many-body states \cite{Kessler2012,Diehl2010,Schwager2013}. Also, a recent proposal for a quantum simulator based on Rydberg atoms relies on dissipation \cite{verstraete2009,Weimer2010}.\\

So far, Rydberg excitations in cold gases have been studied almost exclusively in the resonant, non-dissipative regime. In those experiments, the strong interactions between Rydberg atoms manifest themselves either as spatial correlations compatible with a radius of blockade around an excited atom \cite{Schwarzkopf2011,Schauss2012} or through a reduction of fluctuations leading to sub-Poissonian statistics \cite{Viteau2012}. In this work we show that for off-resonant excitation, the behaviour of the system depends strongly on the detuning and the sign of the Rydberg-Rydberg interaction. We characterize the properties of our system through the full counting statistics of the excitation events, similarly to the methods recently used in condensed matter physics to unveil correlations in electronic transport processes \cite{braggio2006,gogolin2006}. When the detuning and the interaction have the same sign, excitation of pairs of Rydberg atoms can occur, leading to an asymmetry of the excitation lineshape as a function of detuning. In the dissipative regime, we observe strongly bimodal counting distributions in the full counting statistics \cite{Hu2013,Ates2012,Lee2012}, indicating phase coexistence and intermittency in the system \cite{carr2013}.\\

\begin{figure}[h]
\centering
\includegraphics[width=1 \textwidth]{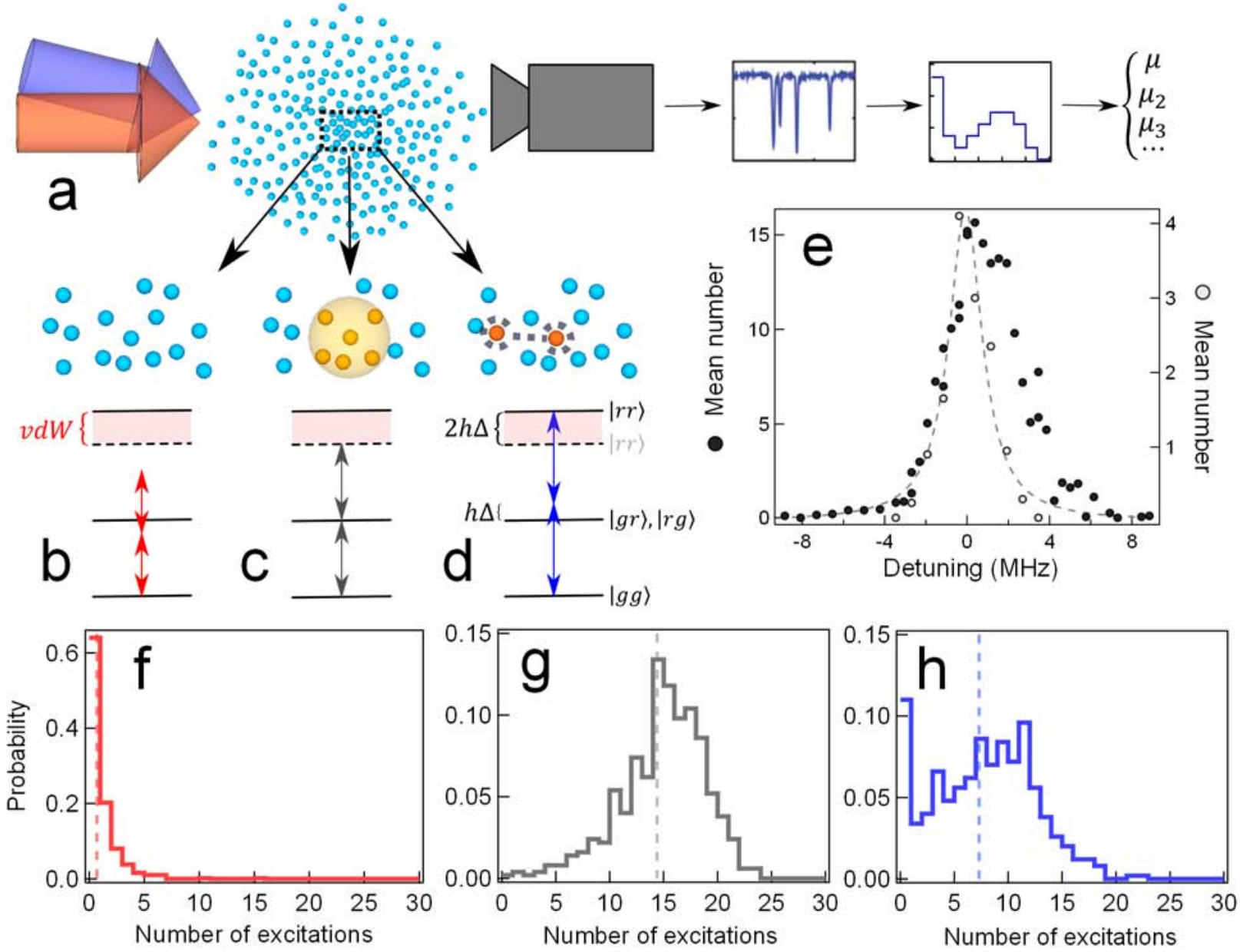}
\caption[1]{Full counting statistics of on- and off-resonant Rydberg excitations. {\bf a}, Schematic representation of the experimental procedure. The cold rubidium atoms are excited to Rydberg states using two laser beams. The resulting Rydberg excitations are detected on a channeltron after field ionization, and from the individual counts the histograms and full counting statistics (i.e., the moments of the counting distribution) are calculated. {\bf b-d}, Resonant and off-resonant excitation processes for interacting Rydberg atoms. For resonant excitation, {\bf c}, the van-der-Waals interaction between the atoms shifts a second excitation out of resonance, leading to the dipole blockade and the associated blockade sphere. Off resonance, interactions between pairs can either lead to resonant pair excitation, {\bf d}, or to strong suppression of excitations, {\bf b}. In {\bf e}, a plot of the mean number of Rydberg excitations as a function of detuning reveals the interaction-induced resonances for positive detuning, which shows up as an asymmetry of the lineshape. Here and throughout the paper we omit the subscript $obs$ indicating the observed quantities (see Methods). The excitation durations are $1\,\mathrm{\mu s}$ (grey symbols, right vertical axis) and $20\,\mathrm{\mu s}$ (black symbols, left vertical axis). {\bf f-h} The histograms of the counting distributions in the resonant and off-resonant regimes reflects the differences in the excitation process.  For positive detuning $\Delta/2\pi=+3.5\,\mathrm{MHz}$, {\bf h}, the histogram exhibits a bimodal structure, whereas on resonance, {\bf g}, it has a single peak. For negative detuning ($\Delta/2\pi=-3.5\,\mathrm{MHz}$), {\bf f}, the mean number of excitations is considerably smaller than in {\bf h}. The dashed vertical lines indicate the mean number of excitations. The Rabi frequency is $2\pi\times400\,\mathrm{kHz}$, the interaction volume $10^{-7}\,\mathrm{cm}^3$ and the density $1.8\times 10^{11}\,\mathrm{cm}^{-3}$.}
\label{stueckel_scheme}
\end{figure}

Figure 1 summarizes the experimental method and typical results of the full counting statistics. Our experiments are performed using 87-rubidium atoms in magneto-optical traps (MOTs) that are excited to $70S$ Rydberg states using a two-step excitation process with detuning $\Delta$ from resonance and two-photon Rabi frequencies of up to $400\,\mathrm{kHz}$ \cite{Viteau2012} (see Methods for details). After the excitation pulse the Rydberg atoms are field ionized and detected using a channeltron charge multiplier, whose output is recorded on an oscilloscope. Three different excitation regimes can be distinguished. Away from resonance, two Rydberg atoms at a distance $r$ from each other can undergo a pair excitation if their van-der-Waals interaction, characterized by the $C_6$ coefficient, matches twice the energy mismatch of the excitation, i.e. $C_6/r^6=2h\Delta$, as shown in Fig. 1d. For the van-der-Waals coefficient of the $70S$ Rydberg state in 87-Rb ($C_6=1.6\,\mathrm{THz}/\mathrm{\mu m^6}$) and a typical detuning of around $10\,\mathrm{MHz}$, this corresponds to a resonant distance $r=6\pm 0.1\,\mathrm{\mu m}$ (where the range in $r$ is due to the finite laser linewidth of about $0.5\,\mathrm{MHz}$). This resonant condition is the exact opposite of the blockade effect (Fig. 1 c), where the interaction leads to the {\em suppression} of excitations, allowing at most one single excitation within a blockade radius (which is collectively shared by all the atoms in the blockade volume). We can estimate the timescale on which off-resonant excitations become important by comparing the resonant collective Rabi frequency $\Omega_{coll}=\sqrt{N_{db}}\Omega$ (where $N_{db}$ is the number of atoms inside a blockade volume) \cite{Comparat2010} with the off-resonant two-photon Rabi frequency $\Omega_{off}=\Omega^2/(2\Delta)$ expected for the two-photon excitation of two Rydberg atoms with an intermediate singly-excited state detuned by $\Delta$ from resonance. For typical values of our experiment ($\Omega\approx 2\pi\times 200\, \mathrm{kHz}$, $N_{db}\approx 50$) we find $\Omega_{coll}\approx 2\pi\times 1.4\, \mathrm{MHz}$ and $\Omega_{off}\approx 2\pi\times 10\, \mathrm{kHz}$ at a detuning $\Delta/2\pi=2\,\mathrm{MHz}$, leading to an off-resonant excitation timescale of around $100\,\mathrm{\mu s}$ (subsequent excitations mediated by already excited atoms can occur on a shorter timescale). Finally, for a detuning with opposite sign to that of the van-der-Waals interaction, neither single-particle nor pair excitations are resonant, leading to a strong overall suppression of the excitation probability (Fig. 1b).\\

These three excitation regimes are summarized in Fig. 1e, where the mean number of Rydberg excitations is plotted as a function of detuning for two different excitation durations. For the $1\,\mathrm{\mu s}$ pulse off-resonant excitations are negligible and hence the lineshape is symmetric. If the pulse duration is increased to $20\,\mathrm{\mu s}$, however, the excitation time is an appreciable fraction of the off-resonant excitation timescale, and off-resonant excitations become visible. This leads to a lineshape that is clearly asymmetric with respect to zero detuning.

Further evidence for the different excitation regimes is found in the full counting statistics of the Rydberg excitations, as represented in Fig. 1f-h by the histograms of the counting distribution. These histograms clearly reveal the qualitative differences between on- and off-resonant excitations: On resonance, the distribution is roughly Poissonian (it becomes sub-Poissonian for even longer excitation durations, for which the system enters the fully blockaded regime, as seen in Fig. 3b). By contrast, for positive detuning the distribution becomes multi-modal with two dominant features: one close to $0$ excitations and the other centered around a mean value of $12$. For negative detuning, the histogram confirms the expected strong suppression of excitations.\\

\begin{figure}[h]
\centering
\includegraphics[width=0.8 \textwidth]{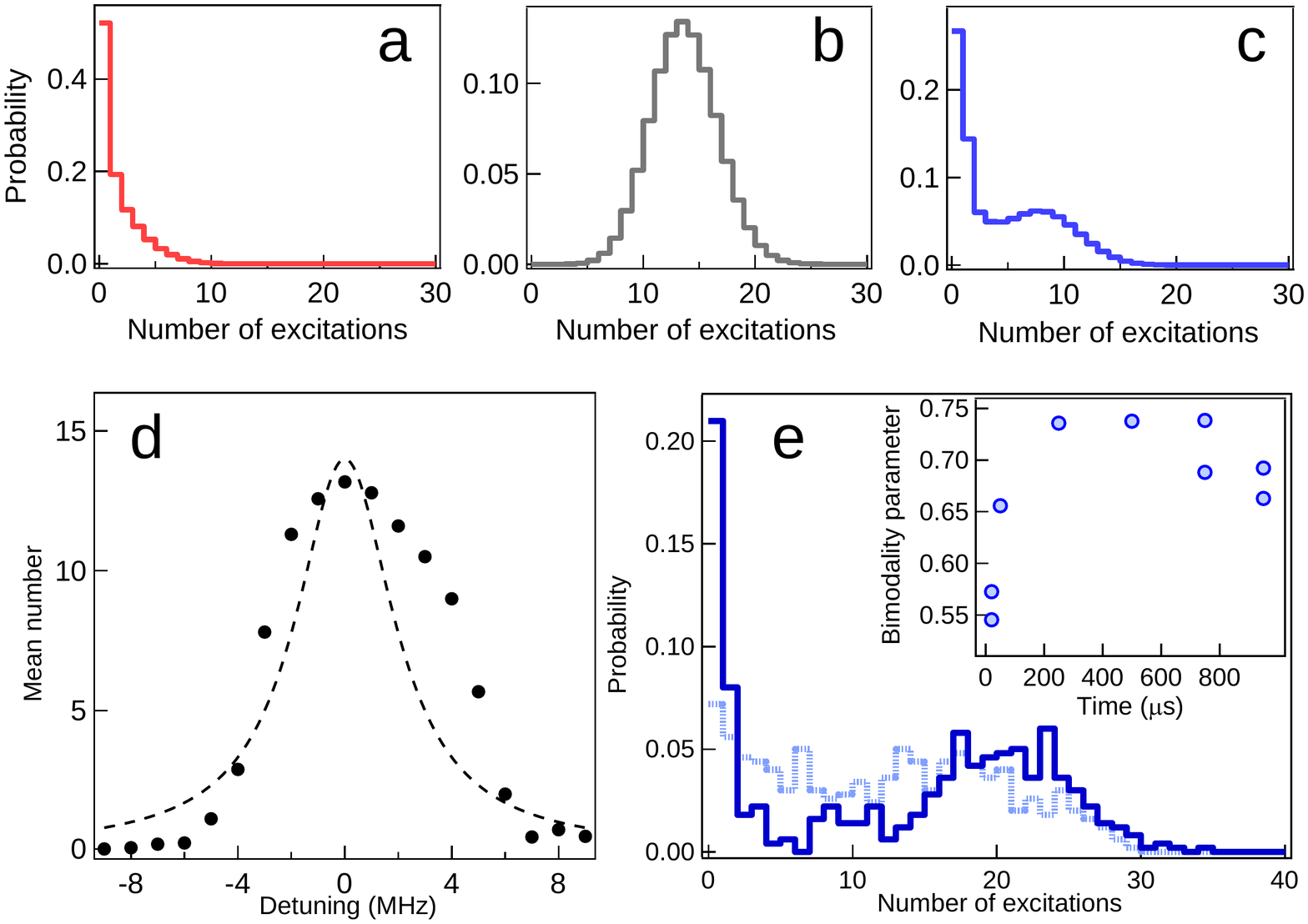}
\caption[1]{Numerical simulations {\bf a-d} and crossover to the dissipative regime {\bf e}. {\bf a-c} show the counting distributions for $\Delta/2\pi=-5,0,5\ \mathrm{MHz}$, respectively, obtained from numerical simulations using the Dicke-type model explained in the Supplementary information. The finite detection efficiency in the experiment is taken into account through a convolution with a binomial function (see Methods). In {\bf d}, the Rydberg mean values obtained from that simulation are plotted as a function of detuning. Comparison with a Lorentzian fit (dashed line) centered at zero detuning highlights the asymmetry of the lineshape. The parameters used for the simulation are: Rabi frequency $130\,\mathrm{kHz}$, excitation duration $4\,\mathrm{\mu s}$, density $9\times 10^{10}\,\mathrm{cm}^{-3}$ and interaction volume $8.9\times 10^{-8}\,\mathrm{cm}^3$.  {\bf e} Experimental results showing the crossover to the dissipative regime. As the duration of the excitation is increased from $20\,\mathrm{\mu s}$ (dashed line) to $950\,\mathrm{\mu s}$ (solid line), keeping the mean Rydberg number fixed by adjusting the Rabi frequency, the counting distributions become more strongly bimodal. The inset shows the bimodality parameter as a function of the excitation duration.}
\label{stueckel_scheme2}
\end{figure}

The main features of the experimentally observed lineshape and counting distributions are reproduced by a numerical simulation based on a Dicke-type model, as explained in the Supplementary information. The main results of that simulation are shown in Fig. 2a-d. Whereas the model is not expected to give quantitative agreement with the experiment (as it does not, for instance, take into account the spatial inhomogeneity of the cloud, the residual velocity of the atoms or decay from the Rydberg state), for reasonable choices of the parameters in the simulation the qualitative agreement with the experimental data of Fig. 1 is rather good.\\

We now turn to the dissipative excitation regime in which several excitation-spontaneous emission cycles occur. Even though our experimental system does not allow us to work in the fully coherent regime (due to limitations regarding the laser linewidth, and hence coherence time, and the maximum Rabi frequency), the $20\,\mathrm{\mu s}$ excitation duration used for the experiments shown in Fig. 1 is an order of magnitude shorter than the lifetime of the $70S$ Rydberg state of around $200\,\mathrm{\mu s}$. A comparison of the counting distributions for different excitation durations (in which the Rabi frequency was adjusted in order to keep the mean value constant, allowing a direct comparison between the histograms), as shown in Fig. 2e, suggests that the counting distribution becomes more strongly bimodal for longer excitation durations, indicating that dissipation favors the appearance of bimodality (as theoretically predicted in \cite{Ates2012,Lee2011,Lee2012}). This is confirmed by the increase in the bimodality parameter (see Methods) as a function of the duration (again for fixed mean Rydberg number).\\

\begin{figure}[h]
\centering
\includegraphics[width=1 \textwidth]{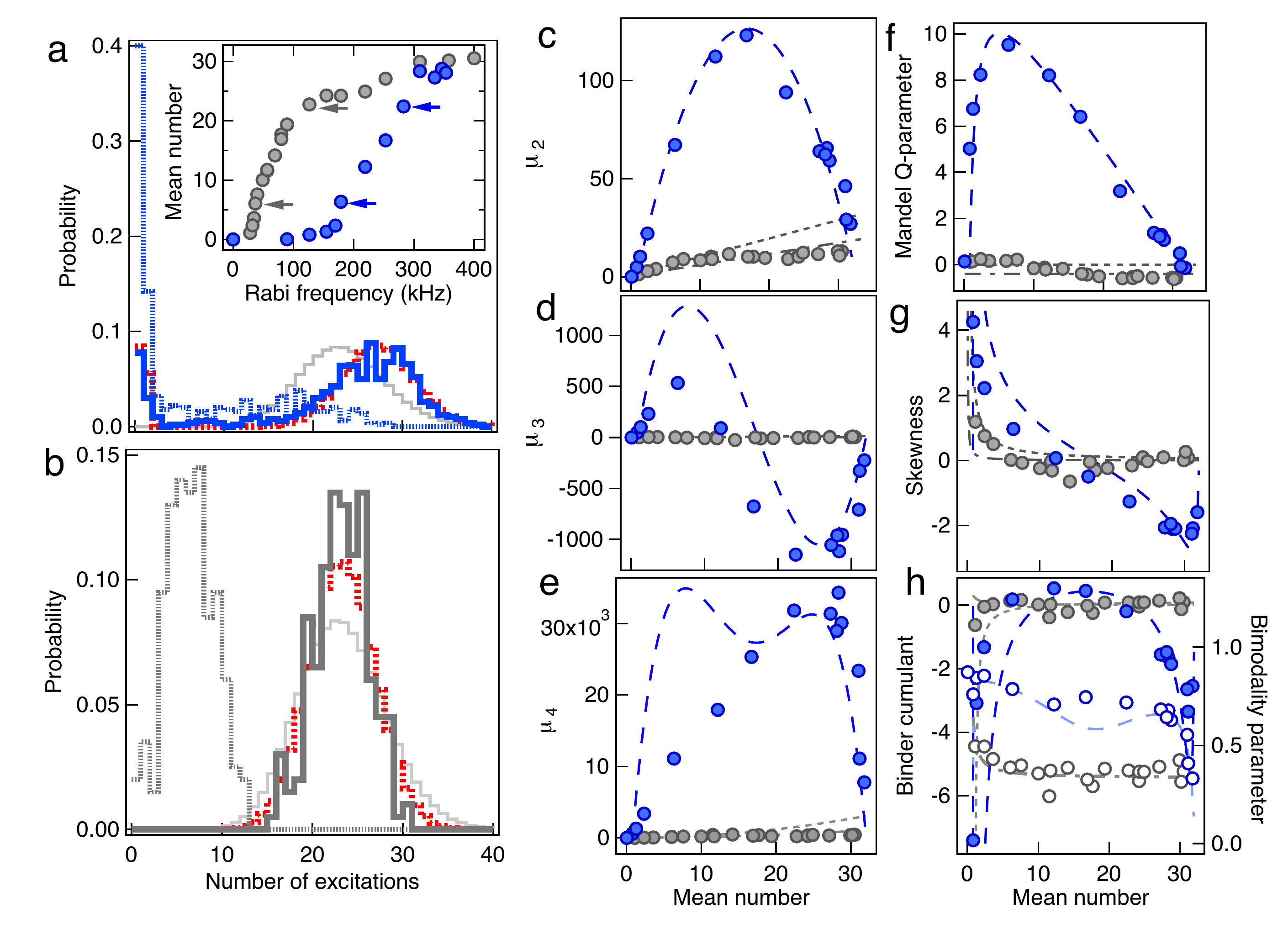}
\caption[1]{Full counting statistics for off-resonant excitation in the dissipative regime. {\bf a} and {\bf b} show off-resonant ($\Delta/2\pi = 11.5\,\mathrm{MHz}$) and resonant counting distributions, for equal mean numbers $6$ (dashed lines) and $23$ (solid lines), respectively (indicated by the arrows in the inset of {\bf a}, which shows the mean number of Rydberg excitations as a function of the Rabi frequency). The dotted red lines are the results of the bimodal model in {\bf a} and of the perfectly monomodal model in {\bf b}, whereas the solid light gray lines show the distributions expected for Poissonian distribution with the same mean values. {\bf c-e} show the second, third and fourth central moments of the off-resonant (blue) and on-resonant (gray) counting distributions. {\bf f-h} show the associated normalized quantities: Mandel $Q$-factor, skewness, and Binder cumulant as well as the bimodality coefficient. The lines in graphs {\bf c-h} are the results of the simple bimodal (dashed), with $N_1=1$ and $N_2=65$ Rydberg excitations (see Methods), Poissonian monomodal (dotted) and perfectly monomodal (dot-dashed) models described in the text. The detection efficiency of $40\%$ was taken into account. For the bimodal model, the theoretical values were scaled by a factor of $\approx 0.5$ in order to facilitate a qualitative comparison with the experimental data. The interaction volume is $3.6\times 10^{-6}\,\mathrm{cm}^3$, the density $1\times10^{11}\,\mathrm{cm}^{-3}$ and the excitation duration $950\,\mathrm{\mu s}$.}
\label{stueckel_scheme}
\end{figure}

Acquiring the complete counting distribution allows us to gain more insight into the properties of our system in the dissipative regime beyond the behaviour of the mean and the variance \cite{tsai98}. We have characterized the full counting statistics on resonance and for positive detuning over a range of Rabi frequencies by analyzing the second, third and fourth central moments $\mu_2$, $\mu_3$ and $\mu_4$ as well as the associated normalized quantities (Mandel $Q$-factor, skewness, Binder cumulant and bimodality coefficient) as a function of the mean number. Fig. 3 illustrates the main results of this analysis. Figs. 3a and 3b show the off- and on-resonant counting distributions for two mean values (the same in both graphs), illustrating the bimodality for off-resonant excitation and the sub-Poissonian character of the distribution on resonance. This difference is clearly seen in the dependence of $\mu_2$ and the Mandel $Q$-factor on the mean number: Whereas for on-resonant excitation those quantities are consistent with Poissonian distributions for small mean numbers, they become increasingly mono-modal (i.e., strongly sub-Poissonian) for larger numbers as the system enters the fully blockaded regime. By contrast, in the off-resonant case $\mu_2$ initially {\em increases} with the mean, reaching a peak at about half the maximum number of excitations (the Mandel $Q$-factor also exhibits a maximum, which is shifted to smaller numbers due to the normalization by the mean). As can be seen in Fig. 3c-h, the results of the off-resonant case are in qualitative agreement with a simple bimodal model. In particular, we find that the bimodality coefficient (Fig. 3h) of the off-resonant counting distributions is consistently higher (at $0.7-0.8$) than for the resonant case (around $0.4$) over the entire range of the mean number, emphasizing the qualitative difference between the two regimes. It is also obvious, however, that while the agreement is good for $\mu_2$, the deviation increases for the higher central moments. This demonstrates the usefulness of the full counting statistics, which allows comparisons with theoretical predictions that are far more sensitive than the mean and standard deviation alone.\\

The results shown in Fig. 3 can be interpreted in terms of a dynamical phase transition between a paramagnetic and an antiferromagnetic phase in a dissipative Ising model with a transverse field, as shown recently in \cite{Ates2012,Lee2012}. Although in those works the atoms are assumed to be arranged in a crystalline structure (created by an optical lattice) with regular spacing, the distance-selective resonance mechanism described above means that in our experiment the Rydberg excitations should arrange themselves in a regular array with a spacing that depends on the detuning, as suggested theoretically in \cite{Garttner2013}. In Fig. 4 we show a phase diagram for our system as a function of the Rabi frequency and the detuning. The mean Rydberg number  as a function of the Rabi frequency exhibits a smooth crossover between $0$ excitations and a maximum number of around $30$ excitations (see also Fig. 3a), where the position of the crossover depends on the detuning. This is expected from the analogy with an Ising spin system \cite{weimer2008}, where the  critical value of the transverse field (which corresponds to the Rabi frequency in our system) increases with increasing Ising interaction (corresponding to the detuning in our case). Moreover, we observe a distinct increase in the Mandel $Q$-factor in the transition region, which is compatible with the intermittent behaviour of the system theoretically predicted in \cite{Ates2012,Lee2012} (and with our observation of bimodal counting statistics in that region), whereby in the transition region the active and inactive phases of the system coexist. In order to prove that interpretation directly, however, it will be necessary to observe the time evolution of a single experimental realization, e.g., through the observation of photons emitted during the decay process \cite{carr2013}.\\

\begin{figure}[h]
\centering
\includegraphics[width=1 \textwidth]{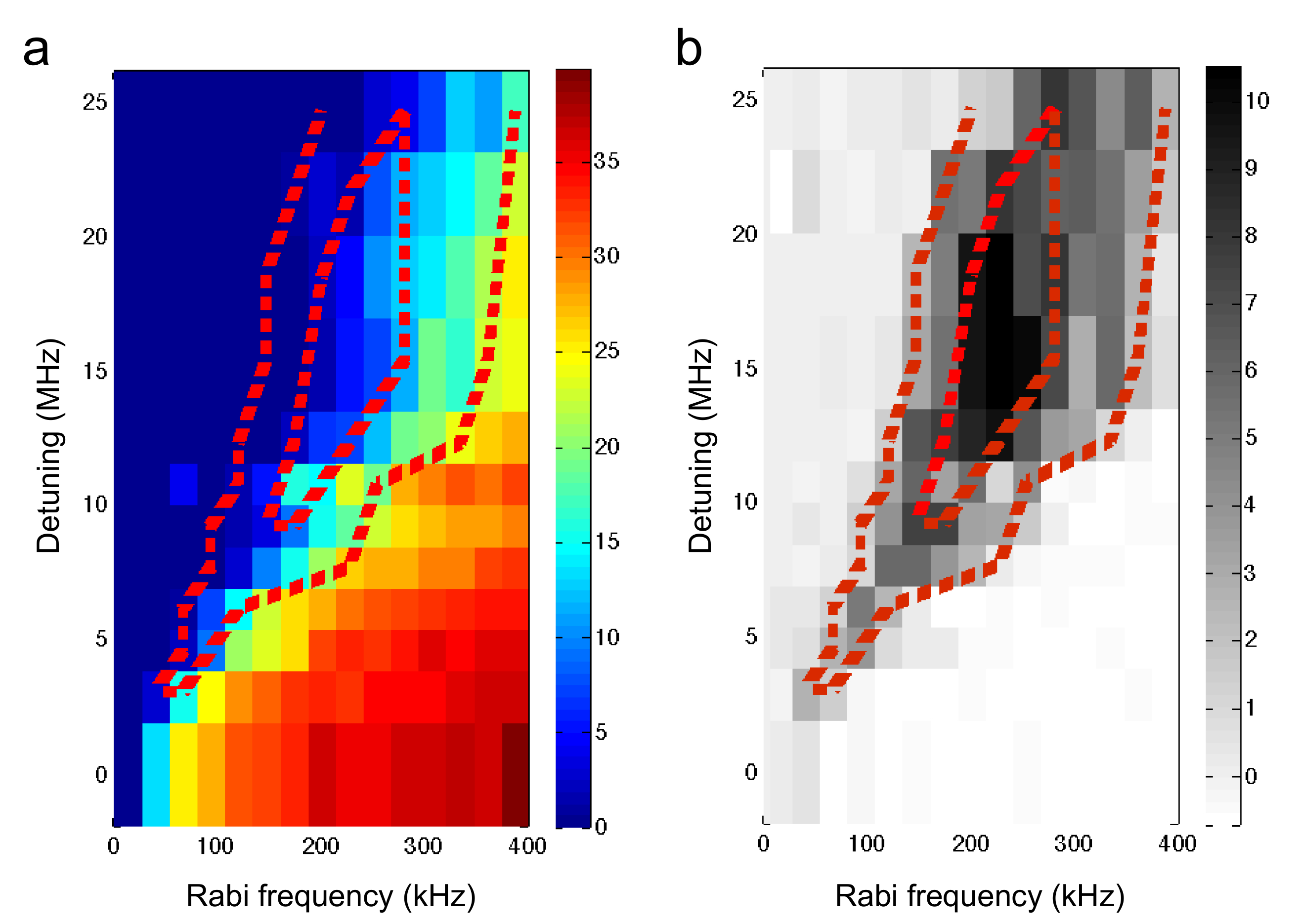}
\caption[1]{Phase diagram in the dissipative regime. {\bf a}, The mean number of Rydberg excitations as a function of Rabi frequency and detuning. {\bf b} The Mandel $Q$-factor as a function of Rabi frequency and detuning. In {\bf a} and {\bf b}, the red dashed lines indicate the transition to $Q>1$ and to $Q>7$. The interaction volume is $3\times 10^{-7}\,\mathrm{cm}^3$ and the density $1.6\times10^{11}\,\mathrm{cm}^{-3}$.}
\label{stueckel_scheme}
\end{figure}

In conclusion, we have analyzed resonant-and off-resonant Rydberg excitations in a cold rubidium gas in the dissipative regime through full counting statistics. We have shown that the full counting statistics reveals characteristic features of the system that are not evident in the mean or standard deviation typically measured in such experiments. Our technique should be useful for the characterization of Rydberg excitations in optical lattices \cite{Viteau2011} in which, e.g., the Ising model can be realized, and in experiments with chirped excitation lasers aimed at the adiabatic creation of Rydberg crystals \cite{pohl2010,Schachenmayer2010,vanbijnen2011}. More generally, the full counting statistics will be an important tool for unveiling many-body effects in Rydberg excitations, for instance in quantum simulators.

\section*{Methods}

\subsection{Experimental setup}
Our experiments are performed using 87-rubidium atoms in magneto-optical traps (MOTs) that are excited to $70S$ Rydberg states using a two-step excitation process with two lasers at 420 nm and 1013 nm, respectively. The first step laser is detuned sufficiently from the intermediate 6P state (by about 2 GHz) to ensure that the intermediate step is not populated on the timescale of the experiment (for our scheme, this could lead to a two-step ionization). The laser at 1020 nm is focused to a waist of 80 microns, and the waist of the 420 nm laser can be varied between 6 and 40 microns, allowing us to change the overlap between the atomic clouds of size $\approx 30-100\,\mathrm{\mu m}$ and the excitation lasers, and hence the effective excitation volume, which can be varied between $\approx 10^{-7}\,\mathrm{cm}^3$ and $\approx 5\times 10^{-6}\,\mathrm{cm}^3$. We note here that for the smallest waist of the 420 nm laser, the resulting excitation geometry is quasi one-dimensional since the waist of the laser is then smaller than the dipole blockade radius. The atomic clouds have peak densities around $0.9-5\times 10^{11}\,\mathrm{cm}^{-3}$. After an excitation pulse of duration between 0.5 and 950 microseconds, during which the MOT beams are switched off, the Rydberg atoms are field ionized by applying a voltage to electrodes located outside the vacuum chamber, and the resulting ions are accelerated towards a channeltron charge multiplier, whose output is recorded on an oscilloscope. The overall detection efficiency $\eta$ for the ions was determined to be around 40 percent.

\subsection{Data analysis}
We extract information on the Rydberg excitations present after a single excitation sequence by recording the channeltron signal due to the arrival of single ions on a digital oscilloscope and then using a peak-finding routine in order to determine the number of ions. This procedure is repeated up to $500$ times, and from the resulting counting distributions the mean $\mu$ as well as the $n$-th order central moments $\mu_n=\langle (x-\mu)^n \rangle$ (up to $n=4$) are calculated. From those moments, we also calculate the normalized quantities characterizing the distributions: Mandel $Q$-factor $Q=\mu_2/\mu-1$, skewness $\gamma=\mu_3/\mu_2^{3/2}$, Binder cumulant $B=1-\mu_4/(3\mu_2^2)$ and bimodality coefficient $b=(\gamma^2+1)/(\mu_4/\mu_2^2)$.

We compare our experimental results to three simple models in order to highlight the main features of the dependence of the various quantities on $\mu$: a perfectly monomodal and a Poissonian monomodal model with mean $\mu_{mono}$ and central moments $\mu_{2,3,4,...}=0$ for the perfectly monomodal model and  $\mu=\mu_{mono}$ , $\mu_2=\mu$ , $\mu_3=\mu$ , and $\mu_4=\mu(1+3\mu)$ for the Poissonian model, as well as a bimodal model with two modes at $N_1$ and $N_2$ and probabilities $1-\alpha$ and $\alpha$, respectively. The mean and the central moments for the bimodal model are $\mu=(1-\alpha)N_1+\alpha N_2$ and $\mu_n=(1-\alpha)(N_1-\mu)^n+\alpha(N_2-\mu)^n$.\\

The finite detection efficiency $\eta$ in the experiment, which transforms the actual counting distribution $P(n)$ into an observed counting distribution $(P(n))_{obs}=\sum_{m=n}^\infty \binom{m}{n}\eta^n(1-\eta)^{m-n}P(m)$, is taken into account by using the analytical result $(\mu_n^F)_{obs}=\eta^n \mu_n^F$ for the $n$-th factorial moments \cite{abate76} and the relationship between the factorial and central moments given by \cite{balakrishnan98}
\begin{equation}
\mu_n=(-\mu)^n+\sum_{i=0}^{n-1} \sum_{j=1}^{n-i}\left[(-1)^i  \binom{n}{i} S(n-i,j)  \right]\mu^i \mu_j^F,
\end{equation}
where $S(i,j)$ is the Stirling number of the second kind.
The observed mean and the second, third and fourth central moments are then related to the actual moments by
\begin{eqnarray}
(\mu)_{obs}&=&\eta\mu \nonumber \\
(\mu_2)_{obs}&=&\eta^2\mu_2+(\eta-\eta^2)\mu \nonumber \\
(\mu_3)_{obs}&=&\eta^3\mu_3+3(\eta^2-\eta^3)\mu_2+(2\eta^3-3\eta^2+\eta)\mu \nonumber \\
(\mu_4)_{obs}&=&\eta^4\mu_4+6(\eta^3-\eta^4)\mu_3+[6(\eta^3-\eta^4)\mu+ \nonumber \\ && +(11\eta^4-18\eta^3+7\eta^2)]\mu_2+(3\eta^4-6\eta^3+3\eta^2)\mu^2+ \nonumber \\&&+(-6\eta^4+12\eta^3-7\eta^2+\eta)\mu
\end{eqnarray}
from which the observed values of the normalized quantities can be calculated (for the Mandel $Q$-factor, the particularly simple result $(Q)_{obs}=\eta Q$ holds). Alternatively, the above procedure can be used to calculate the actual moments and normalized quantities from the observed ones.\\

\subsection{Dicke model}
The numerical simulations shown in Fig. 2 are based on an extension of the Dicke model applied to Rydberg excitations~\cite{Viteau2012,Stanojevic:2012} which is explained in detail in the Supplementary information. Briefly, the original Dicke model, based on the laser excitation conserving the cooperative character of the collective atomic wavefunctions, was modified by including the van-der-Waals interactions between the collective Dicke states, taking into account the $R^{-3}$ contribution at short distances for $S$ states. The collective Dicke states contain the full statistical information about the collective Rydberg excitation and therefore allow the calculation of all the moments of the excitation statistics. The improved Dicke model used here takes into account the time-dependent coupling between the Dicke states, which becomes increasingly important as the number of Rydberg excitations increases. For large excitation numbers the non-symmetric Dicke states experience energy shifts and decoherence due to the van-der-Waals interactions. These, in turn, lead to the asymmetric lineshapes and the characteristic histograms, featuring bimodal distributions for positive detuning, as shown in Fig. 2(c).

\newpage
\section*{Supplementary information}
\subsection*{Many-body Dicke model}
\bigskip The main elements of the Dicke model describing the van-der-Waals blockaded Rydberg
excitation of $N$ atoms have been exposed in \cite{Viteau2012,Stanojevic:2012}.
Here we will concentrate on the details of the revised model that are relevant for the features observed in the experiment.

The Dicke model is based on the laser excitation conserving the cooperative character of the collective atomic
wavefunctions.\ The description of the system through the collective Dicke
states contains the full statistical information, not only the average
quantities. All the moments of the excitation statistics can be calculated, in
particular the measured full counting distributions of the Rydberg
excitations. In order to describe the experimental results presented in this paper the Dicke
model treatment was improved. As explained in \cite{Viteau2012}, the
introduction of an orthogonal basis for diagonalizing the van der Waals basis
coupling matrix in the restriction of the space of the non-fully symmetrical
Dicke states allow us to treat the evolution of the system with an single
integro-differential equation for each symetrical Dicke state with $j$ Rydberg
excitations
\begin{align*}
i\frac{da_{j}}{dt}  &  =-\Delta ja_{j}+W_{ss}^{\left(  j\right)  }a_{j}%
+\sqrt{\left(  N-j\right)  \left(  j+1\right)  }\frac{\Omega}{2}a_{j+1}%
+\sqrt{\left(  N-j+1\right)  j}\frac{\Omega}{2}a_{j-1}\\
&  -i\int_{0}^{t}f^{\left(  j\right)}\left(  \tau\right)  \exp\left(
i\delta j\tau\right)  a_{j}\left(  t-\tau\right)  d\tau,
\end{align*}
where $a_{j}$ is the amplitude of the Dicke state in the collective wave function,
$\Delta$ is the laser detuning, $\Omega$ is the Rabi frequency for the laser
excitation and $W_{ss}^{\left(  j\right)  }$ is the mean value of the van der
Waals interaction for the atoms in the symmetrical $j$ Dicke state. The term
$W_{ss}^{\left(  j\right)  }a_{j}$ in the equation corresponds to the mean
field seen by the atoms. The term $f^{\left(  j\right)  }\left(  \tau\right)  $\ is a kernel
function defined as%
\begin{equation}
f^{\left(  j\right)}\left(  \tau\right)  =\underset{\left\{  q\right\}  }{\sum}\frac{1}{M_{j}%
}\left[  \left[  -\frac{d^{2}}{d\tau^{2}}-2i\frac{d}{d\tau}W_{ss}%
^{(j)}+\left(  W_{ss}^{(j)}\right)  ^{2}\right]  \exp\left(  -iW_{qq}%
\tau\right)  \right]
\end{equation}
where $\left\{  q\right\}  $ corresponds to the ensemble of the $M_{j}%
=\binom{N}{j}$ van der Waals states with $j$ Rydberg excitation and an energy
$W_{qq}$. The evaluation of the kernel function is a crucial and non-trivial step in solving these equations \cite{Stanojevic:2012}. Nevertheless, in ref.~\cite{Viteau2012} we have shown that, to first approximation, the kernel
function term firstly compensates the mean-field term and secondly acts as a
coupling between the fully symmetrical $j$ Dicke state and one particular
virtual \red{ad-hoc} state with the same number of Rydberg excitations, simulating the coupling with
the ensemble of the non-symmetrical Dicke states. This
previous approach was limited to short interaction times $t\lesssim \Omega^{-1}$ and small numbers of Rydberg excitations $N_{Ry}$. The striking
features of the collective dynamics appear at longer interaction times, when
the organisation of the Rydberg excitations within the ultracold cloud has
reached a stationary state. A more complete treatment needs to take into
account the time-dependent coupling between the different states, which
becomes more and more important when $N_{Ry}$ increases.

The kernel function depends on the distribution of the atoms inside the experimental sample. Analytical solutions can be found for many
configurations, but the exact solution of the equations is quite involved. It is easier, and more
interesting for a physical understanding of the problem, to consider the Fourier transform $F^{(j)}\left(  \omega\right)$ of the above kernel, which for a
homogeneous atomic density is given to a good approximation by%
\begin{equation}
F^{(j)}\left(  \omega\right)  =\left(  \omega_{W_j}-\omega\right)  ^{2}%
\Theta\left(  \omega-\omega_{V_{j\max}}\right)  \Theta\left(  \omega_{V_{j\min}
}-\omega\right)  \frac{\omega_{V_{j\max}}^{1/2}}{2\omega^{3/2}},
\end{equation}
with the relation%
\begin{equation}
\omega_{V_{j\max}}\omega_{V_{j\min}}=\omega_{W_j}^{2}%
\end{equation}
where $V_{j\max}$ is related to the volume of the atomic sample, $V_{j\min}$
corresponds to an effective minimum volume given by the mean distance between two atoms and
$\omega_{Wj}=W_{ss}^{\left(  j\right)}$ is the van der Waals coupling.
This coupling can be estimated by taking into account the
$C_{3}R^{-3}$ contribution at short distances for the $S$ states %
\begin{equation}
\omega_{Wj}=W_{ss}^{\left(  j\right)  }=\frac{j(j-1)}{2}\omega_{W}\simeq
\frac{j(j-1)}{2}\left(  \frac{4\pi}{3}\right)  ^{2}\frac{C_{6}}{VV_{0}}%
\end{equation}
where we introduce an effective volume $V_{0}$ as%
\begin{equation}
\frac{C_{6}}{C_{3}V_{0}} \sim 1.
\end{equation}
$\omega_{Vj\max}<\omega_{Wj}$ corresponds to the minimum energy of $j$ Rydberg
atoms contained in the volume $V$. It needs to be estimated for each $j$.

\begin{figure}[h]
\centering
\includegraphics[width=0.8 \textwidth]{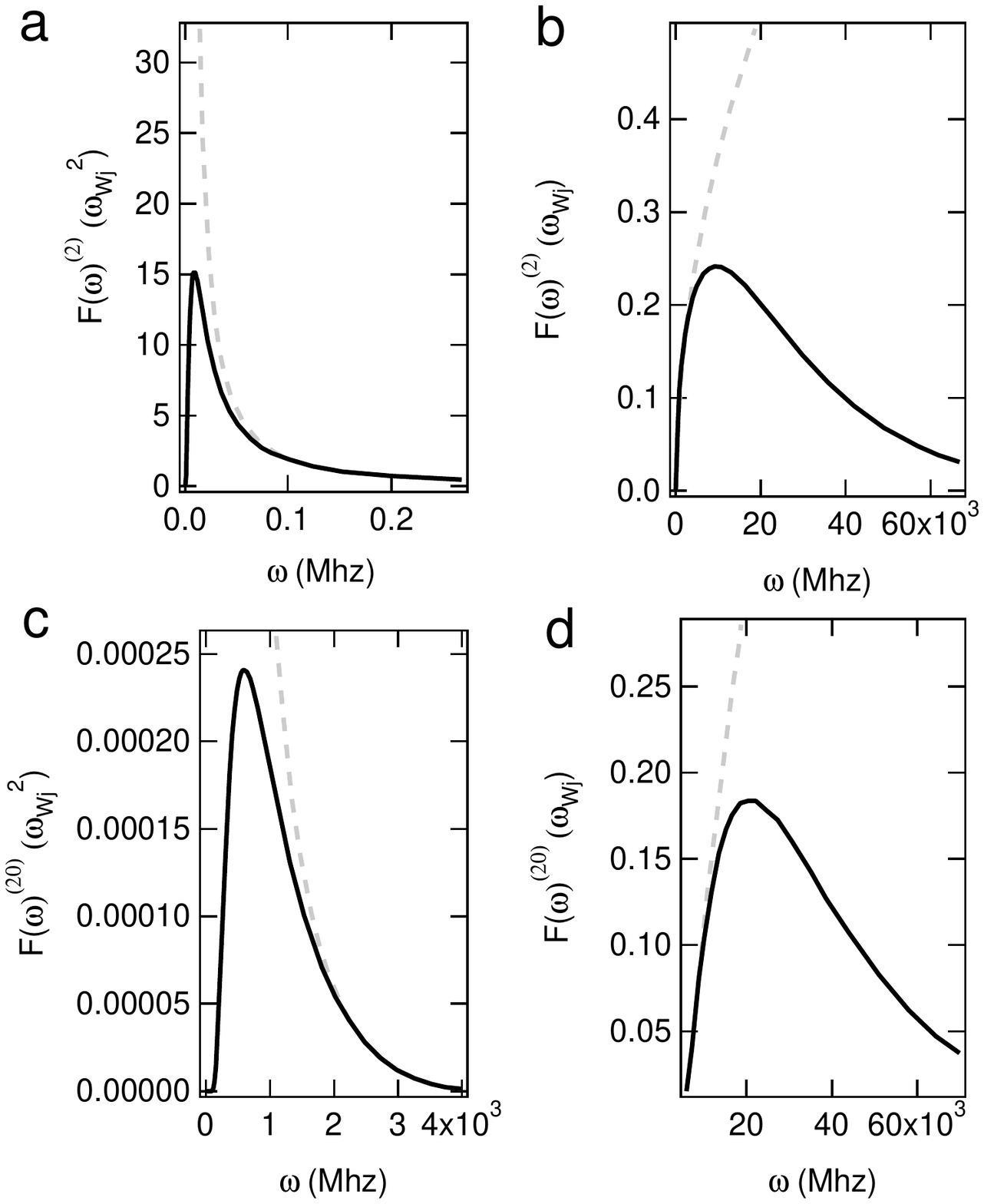}
\caption[1]{Fourier transforms $F^{(j)}\left(  \omega\right)$ of kernel function for $j=2$  {\bf a,b} and $j=20$ {\bf c,d}, calculated at low and high values of $\omega$. Generally, we have a large peak for $\omega>\omega_{Wj}$\ {\bf b,d} and a narrower one for
$\omega<\omega_{Wj}$\ {\bf a,c}. The dashed lines are obtained supposing a homogeneous atomic density, and the continuous lines are based on the atomic density of Eq.\eqref{atdensity}.  We assume a $j(j-1)/2$-dependence of the kernel function \cite{Stanojevic:2012}.}
\end{figure}

The general shape of the Fourier transform function is shown in Fig. 5 for $j=2$ and $j=20$. It is
compared with a function calculated by considering an atomic density given by
the expression%
\begin{equation}
D\left(  V_{x}\right)  =\frac{2}{V\sqrt{\pi}}\exp\left[  2\frac{V_{0}}%
{V}\right]  \exp\left[  -\frac{V_{x}^{2}}{V^{2}}-\frac{V_{0}^{2}}{V_{x}^{2}%
}\right]  ,
\label{atdensity}
\end{equation}
where $V_{x}=\frac{4\pi}{3}R_{x}^{3}$, with $R_{x}$ the distance between
two atoms. The choice of this density function, which is not far from a
homogeneous atomic density in a volume $V$, allows us to pursue the
analytical approach by using the expression%
\[
F^{\left(  j\right)  }\left(  \omega\right)  =\Theta\left(  \omega\right)
\sqrt{\frac{\omega_{Vj\max}}{\pi}}\exp\left[  \frac{\omega_{Vj\max}}%
{\omega_{Wj}}\right]  \exp\left[  -\frac{\omega_{Vj\max}}{\omega}-\frac
{\omega}{\omega_{Vj\min}}\right]  \frac{1}{\omega^{3/2}}%
\]

Both functions exhibit three distinct parts corresponding to very different
frequency ranges. For the high frequencies ($\omega<\omega_{Wj}$), the shape of the function is broad, corresponding to a Markovian short memory process,
which in first approximation compensates the mean-field term.\ For the low
frequencies ($\omega<\omega_{Wj}$), the function is much narrower, which
characterizes a long memory process equivalent to the coupling to a bound
state. \ For the frequencies $\omega\sim\omega_{Wj}$, which corresponds to the
van der Waals coupling value, the function is equal to zero. For large $j$, this
behavior is slightly modified. The mean field term is partly restored, while
the coupling to the the virtual state is decreased in amplitude.\ A decay term
appears in the equations corresponding to a decoherence process for the non
symmetrical Dicke states.

The state of the system corresponding to $j$ Rydberg excitations is described
by two differential equations%
\[
i\frac{da_{j}}{dt}=-\Delta ja_{j}+\omega_{Wj}\left(  1-h_{1}\left(  j\right)
\right)  a_{j}+\sqrt{\left(  N-i\right)  \left(  i+1\right)  }\frac{\Omega}%
{2}a_{j+1}+\sqrt{\left(  N-j+1\right)  j}\frac{\Omega}{2}a_{j-1}+\omega
_{Wj}h_{2}\left(  j\right)  b_{j},
\]
and%
\[
i\frac{db_{j}}{dt}=-\Delta jb_{j}+G\left(  j,t\right)  b_{j}+\omega_{Wj}%
h_{2}\left(  j\right)  a_{j}.
\]
$h_{1}(j)$ and $h_{2}(j)$ are functions such as%
\[
\omega_{Wj}h_{1}(j)=\int_{\omega_{Wj}}^{\infty}-\frac{F^{j}\left(
\omega\right)  }{\omega}d\omega
\]
and%
\[
\left(  \omega_{Wj}h_{2}(j)\right)  ^{2}=\int_{0}^{\omega_{Wj}}F^{j}\left(
\omega\right)  d\omega
\]
$G\left(  j,t\right)  $ is given by%
\[
G\left(  j,t\right)  \sim\frac{1-i}{\sqrt{2}}\frac{\sqrt{\omega_{Vj\max}}%
}{\left(  t-it_{0}\right)  ^{1/2}}%
\]
with $t_{0}\sim\omega_{Wj}^{-1}$.

At this point the dynamics of the system is no longer Hamiltonian. In effect, we
have introduced a decoherence process in the ensemble of the
non-symmetrical states, which leads us to consider a population equation to
conserve the ensemble of the atoms%
\[
\frac{d\rho_{j}}{dt}=-2\operatorname{Im}\left[  G(j,t)\right]  \left\vert
b_{j}\left(  t\right)  \right\vert ^{2}.
\]
The non coherent population with $j$ Rydberg excitations corresponds to a total
van-der-Waals coupling energy of $\sim\hbar\omega_{V\max}$, which
characterizes the global correlation of the system.

The Rydberg excitation of the incoherent and non-symmetrical Dicke states is
treated perturbatively through%
\[
\frac{d\rho_{j}}{dt}=-2\operatorname{Im}\left[  \left[  G(j,t)\right]
\right]  \left\vert b_{j}\left(  t\right)  \right\vert ^{2}-g(j)\rho
_{j}+g\left(  j-1\right)  \rho_{j-1}.
\]

These equations depend on very few adjustable parameters,
essentially the frequencies $\omega_{Wj}$\ and $\omega_{Vj\max}$, for which we
can find reasonable estimates based on physical arguments. In the
case of the $S$ states of rubidium which do not present any degeneracy,
the experimental parameters can be directly used. Compared to the
previous treatment in \cite{Viteau2012} the dynamics of the evolution is taken into account more
carefully and two important new features have been introduced. The first one
is the decoherence in the evolution, and the second one is the appearance of the
correlation corresponding to the energy of the virtual state. The Rydberg
excitation of the coherent non symmetrical state can be taken into account
perturbatively, which does not introduce a fundamental change in the
results of the calculations.

We do not expect our model to perfectly reproduce the experimental results, as several details of the experiment (inhomogeneous density of the cloud, other Rydberg levels, decay from the excited state, finite velocity of the atoms) are not taken into account. Also, in our numerical simulations we found that the dynamics of the system depends on several
parameters, a slight change of which leads to significatively different
results. Furthermore, for long excitation times the results depend sensitively on how we take into account the finite coherence time of the laser excitation. Nevertheless, Fig. 2 in the main text shows that for a reasonable choice of parameters in the simulation, which are not too far from the parameters of the experiment, the results of the simulations agree quite well with the experimental data.
In particular, agreement at the level of the full counting statistics (histograms in Fig. 2 (a-c)) is satisfactory. \ For  positive
detuning the histogram exhibits a bimodal structure, while a single peak is
essentially observed at resonance and very few atoms are excited for
negative detuning.\

Further information is obtained by analyzing the dynamics of the system. \ On resonance the atoms are essentially described by
the incoherent anti-symmetrical states, whereas for negative detuning
only fully symmetrical Dicke states are populated.\ On resonance the evolution
quickly becomes incoherent, and the system is described by a statistical mixture of
states. For negative detuning, by contrast, the system is described by a superposition
of a small number of Dicke states with little excitation. Interestingly, on
resonance the total energy of the different non-coherent antisymmetrical
states is low, meaning the system exhibits strong correlations. The histogram
for positive detuning contains both components, resulting in $60\%$ statistical
mixture of states and $40\%$ coherent superposition of Dicke states.\ We also observe that
for long excitation times the coherent antisymmetrical states are marginally populated
for negative or positive detuning, while they remain present for much longer in the case of
resonant excitation.

\section*{Acknowledgments}
This work was supported by PRIN and the EU Marie Curie ITN COHERENCE. The authors thank R. Fazio, A. Tomadin, M. Dell'Orso and R. Mannella for discussions.

\bibliographystyle{apsrmp}

\end{document}